\documentclass{PoS}

\title{Consequences of a possible jet-star interaction in the inner central parsec of Centaurus A}

\ShortTitle{Consequences of a possible jet-star-interaction in Centaurus A}

\author{\speaker{C.~ M\"uller}$^{a,b}$, M.~Kadler$^{a}$,
  K.~Mannheim$^{a}$, M.~Perucho$^{c,d}$, R.~Ojha$^f$,
  E.~Ros$^{e,c,d}$, R.~Schulz$^{a,b}$, J.~Wilms$^{b}$\\
\llap{$^a$} Institut f\"ur Theoretische Physik und Astrophysik,
Universit\"at W\"urzburg, Germany\\ 
\llap{$^b$} Dr.~Remeis Observatory \& ECAP, 96049 Bamberg, Germany\\ 
\llap{$^c$} Observatori Astron\`omic, Universitat de Val\`encia, Paterna, Val\`encia, Spain\\
\llap{$^d$} Departament d'Astronomia i Astrof\`isica, Universitat de Val\`encia, Burjassot, Val\`encia, Spain\\
\llap{$^e$} Max-Planck-Institut f\"ur Radioastronomie, Auf dem H\"ugel 69, 53121 Bonn, Germany\\
\llap{$^f$} NASA Goddard Space Flight Center, USA\\
E-mail: \email{cornelia.mueller@astro.uni-wuerzburg.de}
}

\abstract{
The jet-counterjet
system of the closest radio-loud active galaxy Centaurus~A (Cen~A) can be
studied with Very Long 
Baseline Interferometry (VLBI) on unprecedented small linear scales of
$\sim\,0.018$\,pc. These high-resolution observations provide essential
information on jet emission and propagation within the inner
parsec of an AGN jet. We present the results of a kinematic study
performed within the framework of the Southern-hemisphere AGN monitoring program
TANAMI. Over 3.5\,years, the evolution of the central-parsec jet
structure of Cen~A was monitored with VLBI. These observations reveal
complex jet dynamics which are well explained by a spine-sheath
structure supported by the downstream acceleration occurring where the
jet becomes optically thin. Both moving and stationary jet features
are tracked. A persistent local minimum in surface brightness suggests
the presence of an obstacle interrupting the jet flow, which can be
explained by the interaction of the jet with a star at a distance of
$\sim\,0.4$\,pc from the central black hole. We briefly discuss possible implications
of such an interaction regarding the expected
neutrino and high-energy emission and the effect on a putative planet.
}
\FullConference{12th European VLBI Network Symposium and Users Meeting,\\
                 7-10 October 2014\\
                 Cagliari, Italy}

\begin{document}
\section{Introduction}
At a distance of 3.8\,Mpc \cite{Harris2010}, Centaurus A (Cen~A) is
the closest radio-loud active galactic nucleus (AGN) exhibiting
powerful radio jets.  Thanks to its proximity it is one of the most
interesting targets to address open questions of AGN jet physics. Very
Long Baseline Interferometry (VLBI) enables us to study the jet at
sub-parsec scales \cite{Mueller2011,Mueller2014}. It is detected up
to $\gamma$-ray energies showing a double-humped broadband spectrum
\cite{Abdo2010_cenacore,HESS2009}. Cen~A is a potential candidate
for ultra-high energy cosmic ray (UHECR) emission
\cite{Kim2013,Auger2007} and furthermore it coincides positionally
with the highest energy PeV neutrino event detected by IceCube
\cite{bigbird}.  Detailed investigations of Cen~A can give a
better understanding of jet formation and emission mechanisms in
extragalactic jets in general.

\section{Results from TANAMI monitoring of Centaurus A}
\begin{figure}[!ht]
    \centering
\includegraphics[width=0.5\textwidth]{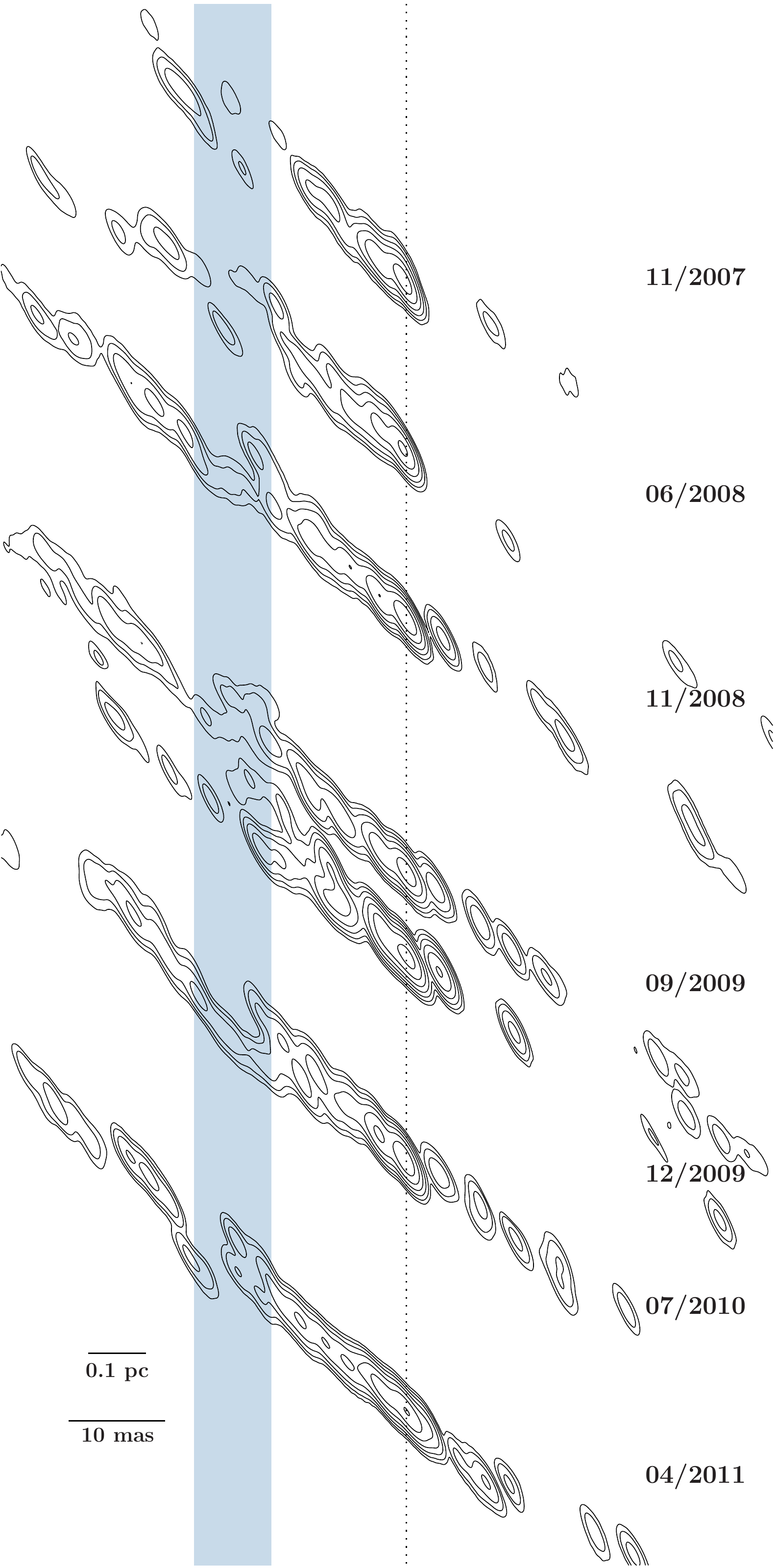}
\caption{
Results from TANAMI monitoring. Shown is the time evolution
  of the inner parsec of Cen~A's jet at 8.4\,GHz. Contour clean images,
restored with a common beam of $(3.33 \times 0.78)$\,mas at
P.A.=$26.3^\circ$, the phase center is indicated by the dashed
vertical line. The contours indicate the flux density level,
scaled logarithmically and increased by a factor of 3, with the lowest
level set to the $5\sigma$-noise-level. The blue shaded area at $\sim\,25$\,mas
away from the phase center highlights the ``tuning-fork'' where the jet
widens and the surface brightness decreases (see \cite{Mueller2014}
for details). This feature can be
explained by a jet-star interaction.}\label{fig1}
\end{figure}

Figure~\ref{fig1} shows the first seven TANAMI (\textit{Tracking
Active Galactic Nuclei with Austral Milliarcsecond Interferometry},
\cite{Ojha2010}) VLBI epochs at 8.4\,GHz of Cen~A \cite{Mueller2014}
using Southern telescopes in Antarctica, Australia, Chile, South
Africa, and New Zealand. With sub-milliarcsecond (mas) resolution we
can study the innermost part of Cen~A's jet in great detail, resolving
individual features at linear scales of $\sim\,0.018$\,pc
(corresponding to 1\,mas at 3.8\,Mpc).
The TANAMI monitoring reveals complex jet dynamics in the inner
parsec. We detect a position-dependent acceleration downstream of the
jet.  The region with higher apparent component speeds coincides with
the part where the jet becomes optically thin \cite{Mueller2011} and
can hence be explained with a spine-sheath structure.  We found a
range of speeds of $0.1c$ to $0.3c$. Using the ratio of the
jet-to-counterjet brightness, the jet angle to the line of sight can
be constrained to $\theta\sim 12^\circ-45^\circ$. This result is
consistent with findings by Hardcastle et al.~\cite{Hardcastle2003}
for the kpc-scale jet, while only our lower limit overlaps with
previous VLBI measurements by Tingay et al.~\cite{Tingay1998} at
$\sim(3-15)$\,mas resolution.
The monitoring also shows long-term stable features. The second
brightest component at $\sim\,3.5$\,mas next to the core is found to be
stationary with almost constant brightness temperature. It can be
identified with the stationary component at $\sim\,4$\,mas found by
\cite{Tingay2001} at lower angular resolution. It has a steep spectrum
\cite{Mueller2011} and can be explained as a jet nozzle locally enhancing
the internal pressure.
Farther downstream, at $\sim\,25$\,mas from the core, the jet shows a
widening and the surface brightness decreases, resembling a ``tuning
fork'' (see Fig.~\ref{fig1}). 
This persistent local minimum in surface brightness suggests
the presence of an obstacle interrupting the jet flow.
Detailed considerations (see \cite{Mueller2014}) lead to a possible
interpretation as the interaction of the jet with the
stellar wind of a red giant.

\section{Jet-star interaction}
When interpreting the ``tuning fork'' observed in the inner parsec of
Cen~A's jet as the interaction with a red giant, further possible
consequences of such an event can be considered. In the following we
point these out and briefly discuss some implications.

\subsection{Production of $\gamma$-rays}
From theoretical calculations and simulations
(e.g., \cite{Araudo2013,Khangulyan2013,BoschRamon2012,Barkov2010}) we
can expect $\gamma$-ray production from the interaction of an AGN jet
with a massive object like a star, though those emission events are on
short timescales (hours to days).  However, assuming multiple interactions within the
jet\footnote{Note, that the radio and X-ray knots in the kpc-scale jet
could be explained by interaction of stars with the jet
\cite{Hardcastle2003}.}, the persistent $\gamma$-ray emission
observed from Cen~A \cite{Abdo2010_cenacore} could be partly
produced by such processes.
Non-thermal particles are generated in the bow shock between the jet
and the stellar wind, leading to synchrotron and inverse Compton
emission. Potential $\gamma$-ray emission from a single jet-star
interaction is the highest for objects close ($\sim\,1$\,pc) to the
jet base \cite{Araudo2013}. According to the calculations by Araudo et
al.~\cite{Araudo2013}, one star
interacting at a distance of $\sim\,1$\,pc with a jet with a luminosity
of $\sim\,10^{43-44}\,\mathrm{erg\,s^{-1}}$ produces a $\gamma$-ray
luminosity of $\sim\,10^{37-38}\,\mathrm{erg\,s^{-1}}$. These values
are comparable to what we can expect from the scenario in Cen~A.
This emission is likely to be outshined by the mildly beamed
$\gamma$-ray emission of Cen~A ($\sim\,10^{41}\,\mathrm{erg\,s^{-1}}$,
\cite{Abdo2010_misaligned}), although such misaligned sources are
potential candidates to be detected at high energies owing to such an event
\cite{Araudo2013}. Furthermore, Barkov et al.~\cite{Barkov2010}
argue that a very high energy (VHE) flare from jet-star interactions
could be detectable, though
not ruling out that such interactions might also contribute to the
persistent VHE emission in Cen~A \cite{HESS2009}.

Intriguingly, the analysis of four years of
\textsl{Fermi}/LAT $\gamma$-ray data of the Cen~A core emission
reveals evidence for a second, harder power law component above
$\sim\,4$\,GeV \cite{Sahakyan2013} which could be due to an
additional emission mechanism besides the jet-related synchrotron
emission.

\subsection{Production of neutrinos}
Prompted by the IceCube detection of ``Big Bird'', one of the three
PeV neutrino events \cite{bigbird}, which is positionally consistent with
Cen~A, we estimate the neutrino events from a jet-star interaction in
Cen~A.
The interaction of protons accelerated at a bow shock with radius
$r_{\rm BS}=3\times 10^{16}$~cm surrounding a red giant and its wind
can also lead to the emission of high-energy neutrinos resulting from
the hadronic production of pions.  In an optically thick target, this
mechanism turns about half of the energy carried by relativistic
protons and ions into neutrinos, and the other half into $\gamma$-rays
\cite{MS97}.
Assuming a red supergiant with a very large radius $r_{\rm RG}=8$~AU
such as UY Scuti, we thus
obtain a maximum neutrino luminosity of
\begin{equation}
L_\nu = {1\over 2} L_{pp\rightarrow\pi}={1\over 2}L_{\rm p}\left(r_{\rm RG}\over r_{\rm BS}\right)^2=
{1\over 2}\left(r_{\rm BS}\over r_{\rm jet}\right)^2 L_{\rm jet}\left(r_{\rm RG}\over r_{\rm BS}\right)^2\approx
5\times  10^{36}~{\rm erg~s^{-1}}
\end{equation}
adopting $L_{\rm jet}=6.5\times 10^{43}$~erg~s$^{-1}$
\cite{{Abdo2010_cenacore}}.  Although the degenerate core of such a
star is optically thick also to the high-energy neutrinos, most of its
cross sectional area is still transparent.

Additional neutrino flux is generated by the optically thin
comet-tail-like wind with $\tau_{\rm pp\rightarrow\pi}= (n_{\rm
p}/10^6~{\rm cm}^{-3})\sigma_{\rm pp\rightarrow\pi}(r_{\rm w}/200~{\rm
AU})\approx 1.4\times 10^{-4}$.  The assumed density of
$10^6$~cm$^{-3}$ corresponds to the expected value for a cylindrical
expansion of the wind from the stellar surface with $n_{\rm
p}=10^9$~cm$^{-3}$ out to 200~AU.
The corresponding neutrino luminosity is given by
\begin{equation}
L_\nu={1\over 2}\tau_{\rm pp\rightarrow\pi}L_{\rm p}\left(r_{\rm w}\over r_{\rm BS}\right)^2=
{1\over 2}\tau_{\rm pp\rightarrow\pi}L_{\rm jet}\left(r_{\rm BS}\over r_{\rm jet} \right)^2\left(r_{\rm w}\over r_{\rm BS}\right)^2
\approx 5\times 10^{35}~{\rm erg~s^{-1}}
\end{equation}
 The Larmor radius of protons $r_{\rm L}\approx 10^{13}(B/6~{\rm
G})^{-1}(\gamma_{\rm p}/2\times 10^7)$~cm is small enough to permit
Fermi acceleration up to multiple PeV energies at the bow shock.
Due to the quasi-stationary target medium and shock, relativistic
beaming does not increase the apparent flux, unless the accelerated
particle distribution is different from the one predicted by the Fermi
mechanism (e.g., beam-like).
Assuming the above neutrino luminosity is concentrated at PeV energies, the expected number of contained PeV neutrino
events during three years of IceCube data taking reaches a value of only  $\sim 10^{-5}$, rendering an association of the Big
Bird neutrino event with a neutrino from a single jet-star interaction with a red giant highly improbable.

\subsection{What would happen to an Earth-like planet?}
In addition to the TANAMI results, the study by Hardcastle et
al.~\cite{Hardcastle2003} also suggests the presence of multiple
jet-star interactions in Cen~A.  It can be assumed that the stars are
at different evolution stages.  The case of a sun-like star orbited by
an Earth-like planet gives rise to the question of how such a planet
would be affected by the interaction.
According to our calculations in \cite{Mueller2014} the
interaction of Cen~A's jet and a sun-like star with a stellar wind of
$v_W=400\,\mathrm{km\,s^{-1}}$ and a mass loss rate of
$\dot{M}=2.5\times 10^{-14}\,\mathrm{M_\odot\,yr^{-1}}$ results in a
distance to the contact discontinuity of $\sim 7$\,AU. Therefore, an
Earth-like planet would still ``be safe'' behind the
contact discontinuity, i.e., not affected by dynamical
effects. For a higher Lorentz factor, the contact discontinuity can reach 1\,AU.
However, we caution that a more detailed statement on the propagation
of the shock front and how it affects the planet requires more
sophisticated simulations taking the planet's atmosphere and
surface conditions into account.

Nevertheless, a lot more life threatening is the $\gamma$-ray emission to
which the planet would be exposed.  Strong $\gamma$-ray radiation can
destroy a putative, protective ozone layer in an Earth-like planet
with harmful consequences for life because of the subsequent increase
of UV radiation reaching the surface (see e.g.,
\cite{Ruderman1974,Thomas2005}).  Thomas et al.~\cite{Thomas2005}
performed detailed calculations of the influence of $\gamma$-ray
bursts (GRBs, with typical luminosities of
$\sim\,10^{51}\,\mathrm{erg\,s^{-1}}$) on the Earth's
atmosphere. According to them, a $\gamma$-ray fluence of
$100\,\mathrm{kJ\,m^{-2}}$ (burst duration 10\,s) would cause significant damage to life due
to $-38$\% ozone depletion (global average). For comparison, $10\,\mathrm{kJ\,m^{-2}}$
($-16$\% depletion) would have a smaller effect and $1000\,\mathrm{kJ\,m^{-2}}$
(up to $-80$\% local depletion) will result in a total damage.
The authors argue that the ionizing X-ray fluence from the burst
afterglow could cause further ozone depletion and a significant amount
of X-ray radiation would arrive at ground without the protective ozone layer. 
Naively adopting the considerations for GRBs by Thomas et
al.~\cite{Thomas2005} and Piran~\&~Jim\'enez~\cite{Piran2014} we can
make simple estimations on the expected $\gamma$-ray fluence at the
position of the
planet while it is exposed to the emission from Cen~A's jet and
compare the result with GRBs.
Taking Cen~A's core $\gamma$-ray luminosity of the order of
$\sim\,10^{41}\,\mathrm{erg\,s^{-1}}$ \cite{Abdo2010_cenacore} and the distance to the AGN core
of a few parsecs\footnote{For this simple
  estimate at these scales the lower
  $\gamma$-ray flux density from the kpc-scale lobes
  \cite{cenalobes2010} can be neglected.}, the ozone layer will be depleted within minutes to a
few hours. This duration is substantially reduced when considering
beamed emission.
To stay below the critical fluence value of
$\sim\,100\,\mathrm{kJ\,m^{-2}}$ the system needs to be at least at a
distance of $\mathcal{O}(200\,pc)$ from the AGN core for a jet
crossing time of about $\sim\,20$\,years. These estimations strongly
depend among other things on the inclination of the system and the
orbital parameters of the planet, e.g., the time for which a close
planet is protected from the $\gamma$-ray radiation by its host star.
Considering other types, from hot or cold rocky Mercury-like to
Jupiter- or Neptune-like gaseous, planets, the impact of such a
star-jet interaction further strongly depends on their magnetic
field, the composition of their atmosphere, and the stellar wind.

\section{Conclusion}

We presented our recent results from the Southern-Hemisphere VLBI
monitoring of Cen~A which reveal complex jet kinematics in the inner
parsec. A long-term stable feature, the ``tuning fork'', can be
interpreted as the interaction of the jet with a star of the host
galaxy. We point out several possible consequences of such a
 scenario which require further follow-up analysis and
simulations to test, e.g., whether there exists a high-energy emission
component in the Cen~A spectrum from jet-star interactions.

\acknowledgments
\small
C.M.\ and R.S. wish to thank RadioNet3 for support to attend this meeting.
C.M.~acknowledges the support of the Bundesministerium f\"ur
Wirtschaft und Technologie (BMWi) through Deutsches Zentrum f\"ur
Luft- und Raumfahrt (DLR) grant 50\,OR\,1404.
E.R. was partially supported by the Spanish MINECO project
AYA2012-38491-C02-01 and by the Generalitat Valenciana project 
PROMETEOII/2014/057.
  M.P. is a member of the work team of projects AYA2013-40979-P and
AYA2013-48226-C3-2-P, funded by the Spanish Ministerio de Econom\`ia y
Competitividad (MINECO), and PROMETEOII/2014/069, funded by Generalitat
Valenciana.
We acknowledge the COST MP0905 action `Black Holes in a Violent Universe'.


\end{document}